\documentclass[11pt,twoside]{article}

\usepackage{asp2006}
\usepackage{epsf}
\usepackage{psfig}
\usepackage{lscape}
\usepackage{graphicx}

\begin{document}
\title{Magnetic Field Structures of BL Lac Objects on Decaparsec Scales}   
\author{P\'eter Veres and Denise C. Gabuzda}   
\affil{Department of Physics, University College Cork, Ireland}    

\begin{abstract} 
Relatively few Very Long Baseline Interferometry (VLBI) polarization observations have been carried out at 18 cm. The importance of such observations lies in their ability to reveal information about the jet magnetic ({\bf B}) field structure and the environment of the jet on scales intermediate between those probed by higher-frequency VLBI and connected-element interferometers such as the Very Large Array. We have obtained polarization observations of 34 BL Lac objects with the Very Long Baseline Array (VLBA), at 4 separate wavelengths in the 18-20 cm band. The 18-cm jets typically extend to tens of parsecs. In some cases, the decaparsec jet is a continuation of the jet on smaller scales, while in others, we see appreciable bending. We have constructed Faraday rotation-measure maps and used them to study the jet {\bf B} field structures and distribution of thermal plasma around the jets. The Faraday rotation is typically large at these wavelengths, and knowledge of the rotation-measure distribution is essential to derive the {\bf B} field structures of the jets. The high sensitivity of these observations to Faraday rotation makes them an effective tool for studies of possible interactions between the jets and the media through which they propagate.
\end{abstract}
\section{Introduction}
BL Lac objects are AGN with weak or undetected optical line emission and highly variable intensity and polarization at all measured wavelengths. \\

The images obtained for this project are among the very first $18$ cm VLBA polarization images of AGN \citep{gabuzda2003,pushkarev2005}. 
We can map the polarization further from the active nucleus than has been possible previously, due to the increased prominence of the optically thin jet emission and the increased sensitivity to faint, extended emission at our relatively long wavelengths.\\

Faraday rotation results in the rotation of the plane of polarization when the radiation travels through plasma with charged particles and {\bf B} fields. The rotation is given by $\Delta \chi \propto  \lambda^2 \int N_e B_{||} dl$, i.e., the integral of the line of sight magnetic field, $B_{||}$, times the electron density $N_e$ along the line of sight.\\

 There is mounting evidence that helical {\bf B} fields are present in the jets of BL Lac objects \citep[and references therein]{gabuzda2007}, and quite possibly other AGN as well. We thus expected that we might find signs of helical {\bf B} fields in our 18 cm polarization observations. Bearing in mind that we see the objects projected onto the plane of the sky and that the jet is pointing at a small angle to the line of sight, if the {\bf B} field is indeed helical and has a relatively large pitch angle, we will see the {\bf B} field vectors as perpendicular to the jet, possibly becoming longitudinal toward the jet edges (a {\bf spine-sheath} structure). Another clue we can look for is a {\bf gradient in the rotation measure} perpendicular to the jet's direction, due to the changing line-of-sight component of the helical field \citep[these proceedings]{mahmud2007a}.
\section{Observations and Reduction}
We observed the 34 BL Lac objects in the complete sample defined by \citet{kuhr_schmidt} over 48 hours with the VLBA at wavelengths of $18.0$ cm, $20.1$ cm, $21.0$ cm and $22.1$ cm on 16-17 January 2004. The preliminary calibration, instrumental polarization calibration and imaging were carried out in the NRAO AIPS package using standard techniques. Faraday rotation is typically large at these wavelengths, and must be taken into account if we wish to  correctly calibrate the polarization angles and to derive the intrinsic B-field structures of the sources. 
 We determined the absolute calibration of the polarization position angles at each of our observing wavelengths by comparing the VLA core polarizations of sevaral compact AGN measured 35 days after our VLBA observations with the polarizations in our VLBA images. Before making the Faraday-rotation maps, we subtracted the Faraday rotation occuring in the Milky Way, indicated by integrated $18 - 20$ cm measurements \citep{pushkarev2001} -- this way, the Faraday-rotation map obtained is due to magnetized plasma near the AGN. 
\section{Results to Date}   
Here we present results for several objects to date. In all cases, the observed polarization vectors have been corrected for the total observed Faraday rotation. In all but one instance (0119+115), we find signs of helical jet {\bf B} fields. 
\subsection{0119+115}
\begin{figure}[ht!]
\begin{center}
\hspace{-1.cm}
\includegraphics[angle=-90,scale=0.5]{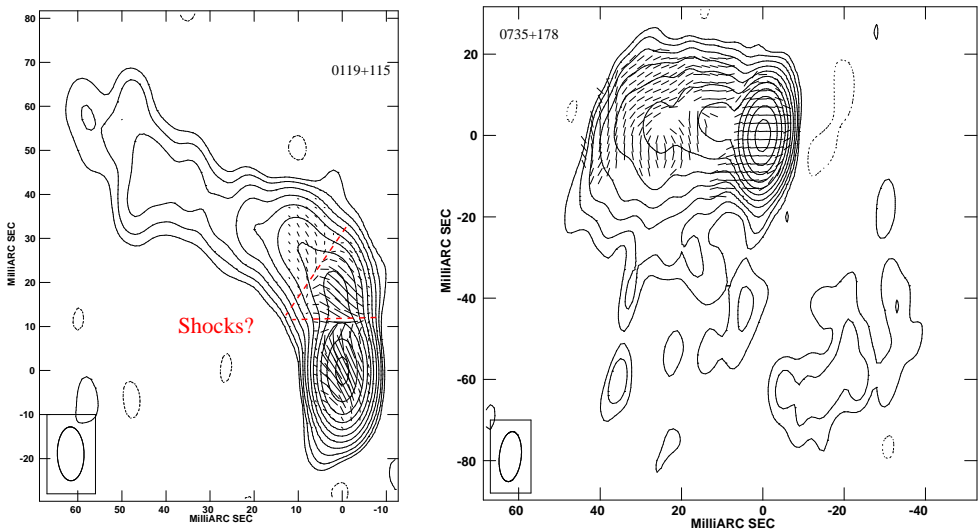}
\vspace{-1.5cm}
\caption{\textit{Left:} {\bf B}-field structure in 0119+115. The marked lines are possible shocked regions. \textit {Right:} The structure of the {\bf B}-field in 0735+178. For clarity, the lengths of the {\bf B}-field sticks have been made identical.}
\vspace{-0.7cm}

\label{fig:jumbo1}
\end{center}
\end{figure}
This compact AGN is unresolved in previous $5$ GHz maps \citep{gabuzda1999}. The overall {\bf B} field is aligned with the jet (Fig. \ref{fig:jumbo1}). Thus, if this jet also has a helical field, the poloidal component is considerably stronger than the toroidal component. The {\bf B} field in this object displays sudden changes in two places (marked in Fig. \ref{fig:jumbo1}), possibly indicative of shocks in the jet that compress the {\bf B} field in a plane perpendicular to the jet direction  \citep[e.g.][]{laing1980,hughes1989}. 
\subsection{0735+178}
In contast to shorter wavelength observations, which show the jet-axis oriented in a roughly NE direction, with a ``zig-zag'' structure visible at some epochs \citep{gomez2001}, the 18 cm jet (Fig. \ref{fig:jumbo1}) extends almost exactly Eastward. There is some evidence that the jet bends toward the South and further to the Southwest, suggesting a spiral-like structure. The superposed {\bf B} field lines indicate that the field is perpendicular to the jet near the central axis, while the field at the northern edge of the jet is oriented along the jet, apparently curving with the jet as it bends -  in other words, we see a spine-sheath structure, characteristic of helical fields. 
\subsection{1803+784}
\begin{figure}[ht!]
\begin{center}
\vspace{-2.5cm}
\includegraphics[scale=0.65,angle=0]{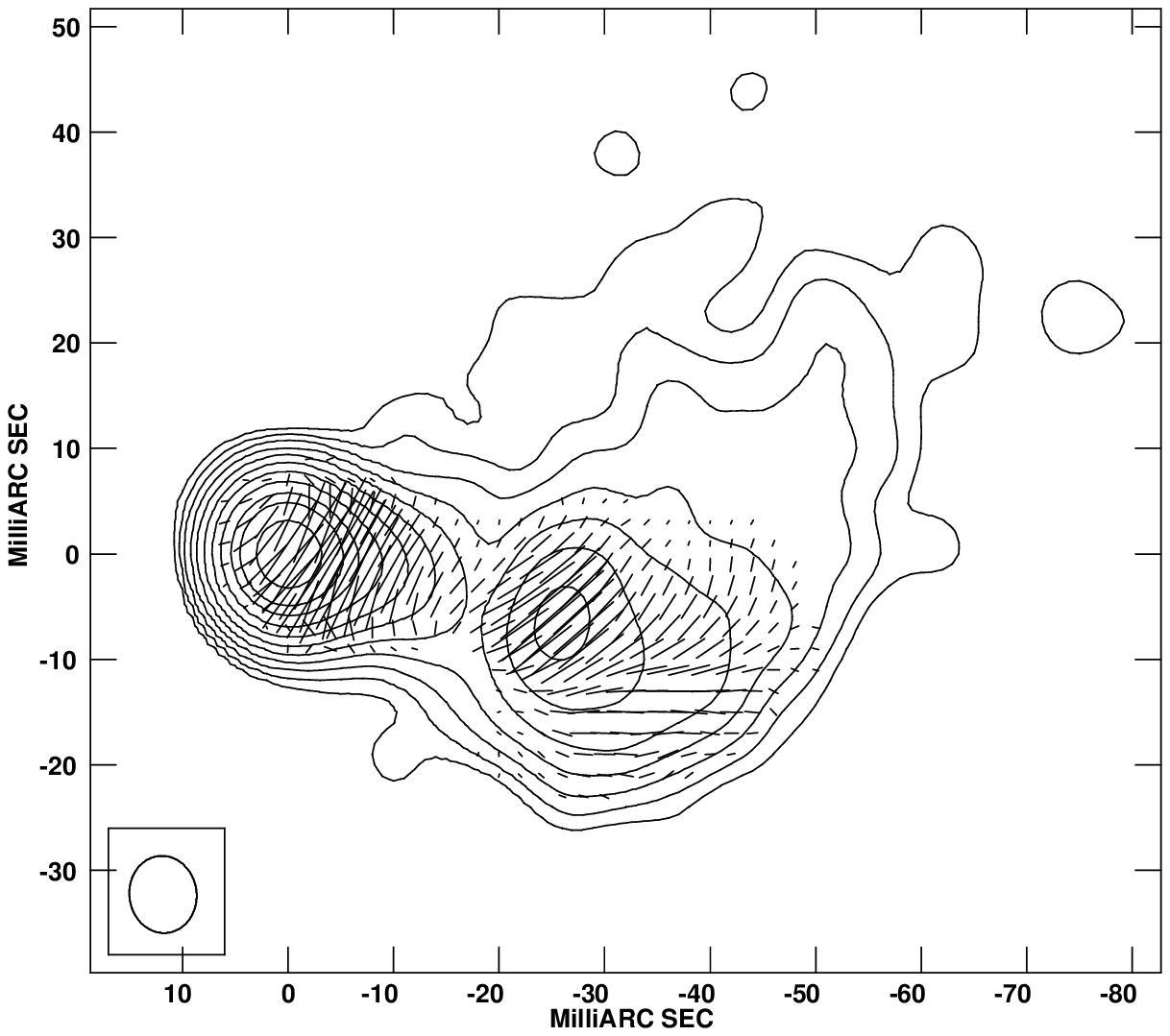}
\vspace{-5.cm}
\caption{\textit {Above:} The direction of the {\bf B} field in 1803+784. \textit {Below:} The Faraday rotation measure in 1803+784 and a slice in the RM distribution perpendicular to the jet direction.}
\label{fig:jumbo2}
\end{center}
\end{figure}

This AGN is known to show evidence for the presence of helical jet {\bf B} fields based on shorter-wavelength measurements \citep[these proceedings]{gabuzda2003,mahmud2007b}. Our 18 cm image (Fig. \ref{fig:jumbo2}) enables us to determine if there is also such evidence further out in the jet.  \\

Our 18 cm image (Fig. \ref{fig:jumbo2}) shows that the {\bf B} field is perpendicular to the jet all along the central part of the jet, suggestive of the toroidal component of a helical field. The {\bf B} field is very different at the southern edge of the jet, and bears no obvious relation to the jet direction. The origin of this {\bf B} field structure is not clear, but may be related to the bend to the north observed in this same region. 
We also find a tentative gradient in the rotation measure across the jet (Fig. \ref{fig:jumbo2}), providing further support for the presence of a helical jet {\bf B} field extending to decaparsec scales in this object. 
\acknowledgements 
This publication has emanated from research conducted with the financial support of Science Foundation Ireland. The National Radio Astronomy Observatory is operated by Associated Universities Inc. 


\end{document}